\documentclass[%
twocolumn,
superscriptaddress,
 amsmath,amssymb,
 aps,
 prl
]{revtex4-2}

\usepackage{graphicx}
\usepackage{dcolumn}
\usepackage{bm}

\begin{document}

\preprint{APS/123-QED}

\title{\textbf{Two-magnon scattering in the framework of the Lippmann-Schwinger equation} 
}%

\author{Jorge Marquez Chavez}
\affiliation{Department of Materials Science and Engineering, Massachusetts Institute of Technology, Cambridge, MA, USA}
\email{jorgemar@mit.edu}

\author{Ondřej Wojewoda}
\email{ondrej.wojewoda@vutbr.cz}
\affiliation{Department of Materials Science and Engineering, Massachusetts Institute of Technology, Cambridge, MA, USA}
\affiliation{CEITEC BUT, Brno University of Technology, Purkyňova 123, 612 00, Brno, Czech Republic}

\author{Yixuan Song}
\affiliation{Department of Materials Science and Engineering, Massachusetts Institute of Technology, Cambridge, MA, USA}

\author{Geoffrey S.D. Beach}
\affiliation{Department of Materials Science and Engineering, Massachusetts Institute of Technology, Cambridge, MA, USA}

\author{Caroline A. Ross}
\affiliation{Department of Materials Science and Engineering, Massachusetts Institute of Technology, Cambridge, MA, USA}

\date{\today}

\begin{abstract}
Controlling magnetic losses requires identifying the microscopic origin of extrinsic linewidth broadening. We introduce a Lippmann--Schwinger framework for two-magnon scattering from weak defect potentials, showing that scattering is set by the overlap between the perturbation of the effective field and degenerate spin-wave states. Applied to iron garnet films, YIG/GGG(111) and YIG/GGG(110), we identify crystallographic directions along which these perturbation are elongated. This approach provides a link between the crystallography, effective field on the sample scale and two-magnon scattering.
\end{abstract}

\maketitle

Efficient information-processing technologies require materials with exceptionally low dissipation \cite{Serga2010, chumak2015magnon, Hirohata2020}. In magnetic systems, losses are commonly described within the Landau–Lifshitz–Gilbert framework through the Gilbert damping parameter \cite{Lakshmanan2011}, which is often extracted from the frequency dependence of the ferromagnetic resonance (FMR) linewidth \cite{Dubs2020}. However, the measured linewidth generally contains additional contributions beyond intrinsic damping, including sample inhomogeneities, two-, three-, and four-magnon scattering processes, as well as field-drag effects \cite{Arias1999, Woltersdorf2004, Lindner2009, Cheng2018, Lacroix2023, Franca2026}. These mechanisms can influence system behavior in fundamentally different ways depending on the specific application, highlighting that reliable device optimization requires accurate interpretation of linewidth data and correct identification of the dominant relaxation process \cite{Yoshii2022}. 

While previous studies have primarily focused on separating linewidth contributions phenomenologically \cite{cheng2018thickness, he2017tunable}, the underlying physical mechanisms governing these processes have not been thoroughly explored \cite{Krivosik2007}. Here, we introduce a framework based on the Lippmann–Schwinger scattering equation and Born approximation for spin-wave scattering from weak defect potentials and reveal the mechanism underlying the in-plane angular dependence of the FMR linewidth \cite{Griffiths2018}. In this approach, defects are treated as localized perturbations of the magnetic effective field that do not significantly modify the spin-wave dispersion. The scattered-wave amplitude can then be expressed as the product of the Fourier image of the effective field contribution from the defect and the density of spin-wave states in reciprocal space (Bloch function) \cite{Wojewoda2024}. Using this framework, we investigate FMR losses in yttrium iron garnet (YIG) films grown on gadolinium gallium garnet (GGG) substrates with (111) and (110) orientations. Comparison of the two substrate orientations shows that the two-magnon scattering strength follows the high-symmetry crystallographic directions. Our results establish a direct link between the perturbed effective field, two-magnon scattering and experimentally observed linewidth anisotropies, which can be exploited in the engineering of future spin-wave-based devices. 

\begin{figure*}
\includegraphics{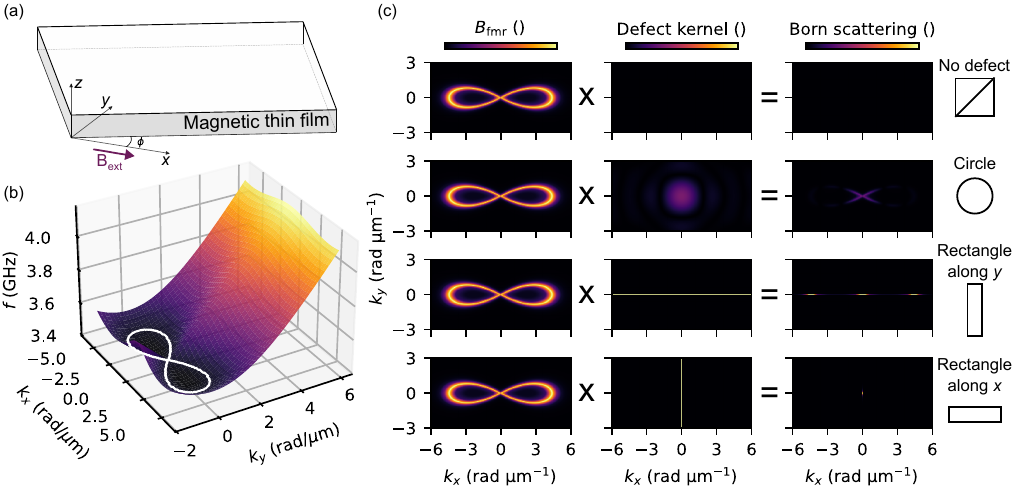}
\caption{\label{fig:Fig1} Schematics of the two-magnon scattering on defects of different geometries. (a) Sketch of the studied geometry. The thin film is in the \textit{xy}-plane and external magnetic field is applied along the \textit{x-}direction. (b) Calculated dispersion relation in two dimensions. The isofrequency line at FMR is marked as a gray solid line. (c) Calculation of the resulting scattering based on the shape of the defect.}
\end{figure*}

In the following text, we will consider a thin ferromagnetic film magnetized by an external magnetic field in the direction of the \textit{x}-axis (Fig.~\ref{fig:Fig1}a). The simulation parameters correspond to those of a YIG film without magnetocrystalline anisotropy and with a thickness of 50\,nm. Two-magnon scattering conserves frequency, requiring the scattered magnon states to lie on the ferromagnetic resonance isofrequency contour in reciprocal space, whose shape is determined by the anisotropic spin-wave dispersion (gray line in Fig.~\ref{fig:Fig1}b). In order to model the efficiency of the scattering and its impact on the linewidth, we will start from the Lippmann–Schwinger scattering equation for magnetization dynamics of a thin film with susceptibility $\hat{\chi}$ and dynamic magnetization vector ($\mathbf{m}$)
\begin{equation}
\begin{aligned}
\mathbf{m}(\mathbf{k},\omega)
&=
\hat{\chi}_0(\mathbf{k},\omega)
\mathbf{h}(\mathbf{k},\omega) \\
&\quad
+ \hat{\chi}_V(\mathbf{k},\omega)
\int \frac{d\mathbf{k}'}{(2\pi)^2}
\, \boldsymbol{V}(\mathbf{k}-\mathbf{k}')
\, \mathbf{m}(\mathbf{k}',\omega).
\end{aligned}
\end{equation}
where $\boldsymbol{V}$ is the scattering potential,  $\mathbf{h}$ is the driving field, and $\chi$ is the susceptibility that can be expressed as
\begin{equation}
\boldsymbol{\chi}_n(\boldsymbol{k},\omega)
=
\frac{1}{\omega-\omega_n(\boldsymbol{k})+i\tau_n(\boldsymbol{k})}
\begin{pmatrix}
|m_{x,n}|^2 & m_{x,n} m_{y,n}^{*} \\
m_{y,n} m_{x,n}^{*} & |m_{y,n}|^2
\end{pmatrix},
\end{equation}
where $\tau$ is the lifetime. Within the Born approximation, the scattering potential is assumed to be sufficiently weak that it does not modify the magnetic susceptibility. The susceptibility inside the defect is therefore taken to be identical to that of the surrounding medium. This allows the expression to be simplified to
\begin{equation}
\begin{aligned}
\mathbf{m}(\mathbf{k},\omega)
&=
\hat{\chi}(\mathbf{k},\omega)
\biggl(
\mathbf{h}(\mathbf{k},\omega) \\
&\quad
+ \int \frac{d\mathbf{k}'}{(2\pi)^2}
\, \boldsymbol{V}(\mathbf{k}-\mathbf{k}')
\, \mathbf{m}(\mathbf{k}',\omega)
\biggr).
\end{aligned}
\label{eq:scatteringBorn}
\end{equation}
Since in FMR the magnetization dynamics is driven by the homogeneous global field in the in-plane direction, we can simplify further by assuming
\begin{equation}
    \mathbf{h}(\mathbf{k},\omega) = \begin{pmatrix}
    \mathbf{h}(\mathbf{k}=0,\omega) \\
    0
\end{pmatrix}
\end{equation}
and writing down only scattered spin-wave states ($\boldsymbol{m}_\mathrm{s}$) yields

\begin{equation}
\mathbf{m}_\mathrm{s}(\mathbf{k},\omega)
\propto
\hat{\chi}(\mathbf{k},\omega)
\, \boldsymbol{V}(\mathbf{k}).
\label{Eq:ScaEff}
\end{equation}

This result shows that the scattering efficiency into individual wavevectors is not only determined by the defect potential, but also by the anisotropic dispersion relation that determines susceptibility $\hat{\chi}$.

In the following discussion, we will simplify the equation by neglecting precessional motion of the magnetization and considering only the scalar spin-wave amplitude, instead of two components of the precession. We calculate the density of the spin-wave states in reciprocal space corresponding to the distribution of Bloch states, as shown in Fig.~\ref{fig:Fig1}c \cite{Wojewoda2024, klima2026, swt_docs}. With no magnetic anisotropy or shape-induced demagnetizing field, this dispersion relation remains the same for any rotation angle of the sample. Now, we will examine different types of  scattering potential in  Eq.~\ref{Eq:ScaEff}. If there is no defect ($V=0$), the two magnon scattering is completely suppressed. For a circular defect, the reciprocal image of scattering potential is isotropic. As a result, the distribution of scattered states follows the same symmetry as the Bloch function, and  rotation of the sample does not lead to a change in the two-magnon scattering strength. The magnitude of the wavevector into which the FMR mode can scatter is determined by the characteristic size of the defect, such that smaller defects enable scattering into higher-wavenumber spin-wave modes.

The scattering from an infinitely elongated rectangular defect is now described for two orientations with respect to the external magnetic field. When the defect is aligned along the \textit{y-}direction (Damon-Eshbach direction), its reciprocal image allows  scattering into  states with wavevector in the \textit{x-}direction (backward volume direction). In this configuration,  scattering on the isofrequency line into the backward volume modes is allowed and we will observe an increase of the linewidth caused by the two-magnon scattering. On the contrary, if the rectangular defect is aligned along the \textit{x-}direction (backward volume),  scattering is allowed into  wavectors along the \textit{y-}direction (Damon-Eshbach direction). Since there are no available states on the isofrequency line, the two-magnon scattering will be suppressed and there would not be any two-magnon contribution to the total linewidth. This behavior was further corroborated by the micromagnetic simulations, see End Matter.

Although the discussion above focused on scattering from the uniform FMR mode, the same framework applies to any spin-wave mode. In this more general case, Eq.~\ref{eq:scatteringBorn} describes elastic scattering into degenerate spin-wave states selected jointly by the mode susceptibility and the reciprocal-space structure of the perturbation. This perspective suggests a route beyond loss analysis: by tailoring the symmetry and length scale of engineered defect potentials, one could deliberately control mode conversion, for example by redirecting propagating spin waves into high wavevector states \cite{Krcma2025} or into caustic spin-wave beams \cite{Wartelle2023, Muralidhar2021}.

\begin{figure}
\centering
\includegraphics[width=\columnwidth]{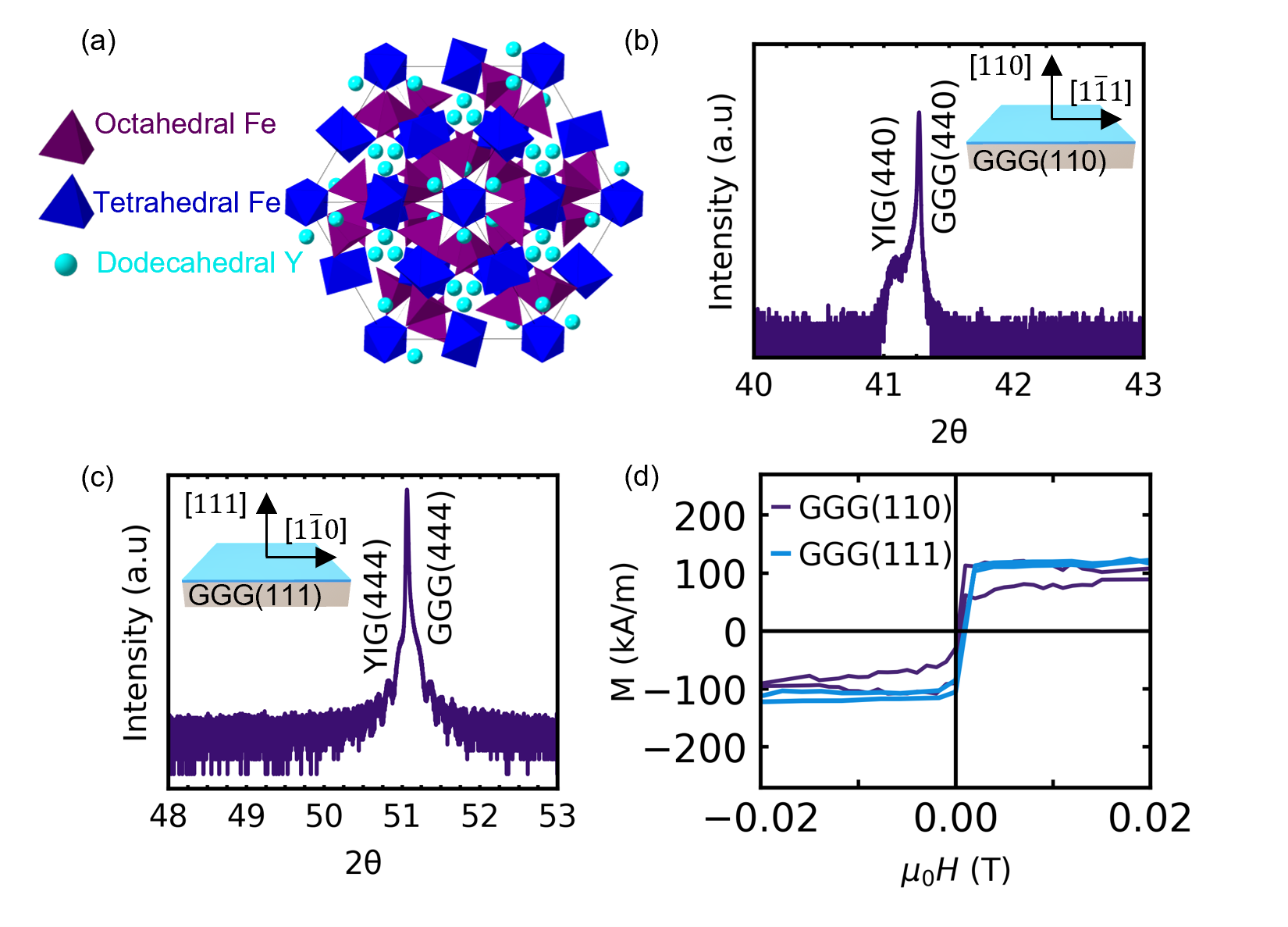}
\caption{\label{fig:Fig2} 
Structural and magnetic characterization of the studied sample.
(a) Garnet crystal structure along <111>, omitting the oxygen atoms.
(b) Symmetric $2\theta$ diffraction scan of 24 nm thick YIG grown on GGG(110) substrate near the (440) reflection.  
(c) Symmetric $2\theta$ diffraction scan of 70 nm thick YIG grown on GGG(111) substrate near the (444) reflection. 
(d) Vibrating sample magnetometry of 70 nm YIG/GGG (110) and 24 nm YIG/GGG(111).}

\end{figure}

To experimentally investigate the angular dependence of magnon scattering, we prepared a 70\,nm thick YIG  film on GGG(111) and a 24\,nm thick YIG film on GGG(110) using pulsed laser deposition (PLD); further details are provided in the End Matter. High-resolution X-ray diffraction (\(2\theta\)) scans near the (440) and (444) reflections are shown in Fig.~\ref{fig:Fig2}b,c, respectively. The YIG(440) peak is observed at lower angles than the corresponding GGG(440) substrate peak, indicating an expanded unit cell volume relative to bulk YIG, which may arise from oxygen deficiency or an  off-stoichiometry of the Y:Fe ratio. In contrast, in the YIG/GGG(111) sample the YIG peak overlaps the substrate peak and Laue fringes are present, consistent with an unstrained film with uniform thickness.  Magnetic hysteresis loops of both samples were measured using vibrating sample magnetometry (VSM) (Fig.~\ref{fig:Fig2}d). The YIG/GGG(110) sample exhibits a saturation magnetization of \(90 \pm 10\,\mathrm{kA/m}\), while the YIG/GGG(111) sample shows a higher value of \(120 \pm 10\,\mathrm{kA/m}\), consistent with previously reported values for YIG films grown by pulsed laser deposition \cite{song2024temperature} but lower than the bulk saturation magnetization of 140\,kA/m. 

\begin{figure}
\centering
\includegraphics{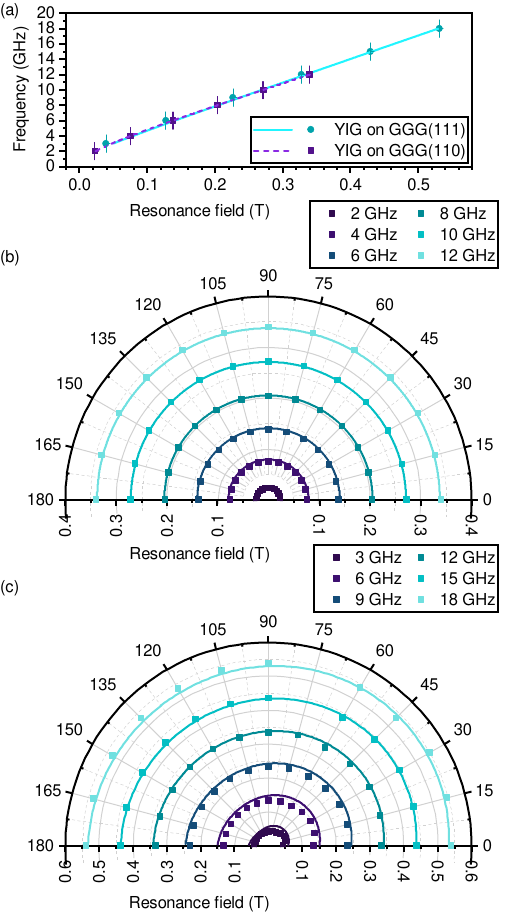}
\caption{\label{fig:Fig3} Ferromagnetic resonance fields for different frequencies.
(a) FMR fields for angle 0\,deg for YIG(111) and 135\,deg for YIG(110).
(b, c) Angle resolved FMR fields for YIG(111) \textbf{(b)} and YIG(110) \textbf{(c)}. }
\end{figure}

Ferromagnetic resonance (FMR) measurements were carried out in a flip-chip geometry, where the sample was mounted face-down on a 50\,$\Omega$ coplanar waveguide (Fig.~\ref{fig:Fig3}a). The samples were subsequently rotated about the out-of-plane axis to determine the in-plane anisotropy (Fig.~\ref{fig:Fig3}b,c). The resonance fields $\mu_{0} H_\mathrm{res}$ were fitted with the Smit-Beljers formula assuming uniaxial anisotropy \cite{Smit1955, puszkarski2012interpretation},

\begin{equation}
\begin{aligned}
\mu_{0} H_\mathrm{res}
&=
\frac{1}{2}
\Biggl\{
-\mu_{0}M_{s}
-
\mu_{0} H_{\mathrm{ani}}
\Bigl[
\cos^{2}\!\bigl(\phi\bigr)
+
\cos\!\bigl(2\phi\bigr)
\Bigr]
\\
&\quad+
\Biggl[
\Biggl\{
\mu_{0}M_{s}
+
\mu_{0} H_{\mathrm{ani}}
\Bigl[
\cos^{2}\!\bigl(\phi\bigr)
+
\cos\!\bigl(2\phi\bigr)
\Bigr]
\Biggr\}^{2}
\\
&\qquad
-
4
\Biggl\{
\mu_{0} H_{\mathrm{ani}}
\cos\!\bigl(2\phi\bigr)
\Bigl[
\mu_{0}M_{s}
+
\mu_{0} H_{\mathrm{ani}}
\cos^{2}\!\bigl(\phi\bigr)
\Bigr]
\\
&\qquad\qquad
-
\left(\frac{\omega}{\gamma}\right)^{2}
\Biggr\}
\Biggr]^{1/2}
\Biggr\},
\end{aligned}
\end{equation}

where $\mu_{0} H_\mathrm{ani}$ is the in-plane anisotropy field, $M_s$ is the saturation magnetization, $\mu_0$ is the vacuum permeability, $\omega$ is the angular frequency, $\phi$ is the angle between the external magnetic field and the easy axis, and $\gamma$ is the gyromagnetic ratio. For YIG(111) the fit showed no in-plane anisotropy, while in the case of YIG(110) it yields an in-plane anisotropy field of $\mu_{0} H_\mathrm{ani, YIG(110)} = 11 \pm 0.2$\,mT in the direction of $\phi_\mathrm{YIG(110)}=131 \pm 6$\,deg. The magnetocrystalline anisotropy in the (110) plane is $K{_1}/4$ which predicts an in-plane anisotropy field of only 2\,mT for bulk YIG. The measured anisotropy likely includes contributions from both magnetoelastic anisotropy of the strained film and growth-induced anisotropy \cite{kaczmarek2024atomic} resulting from the non-ideal Y:Fe ratio.

\begin{figure*}
\centering
\includegraphics{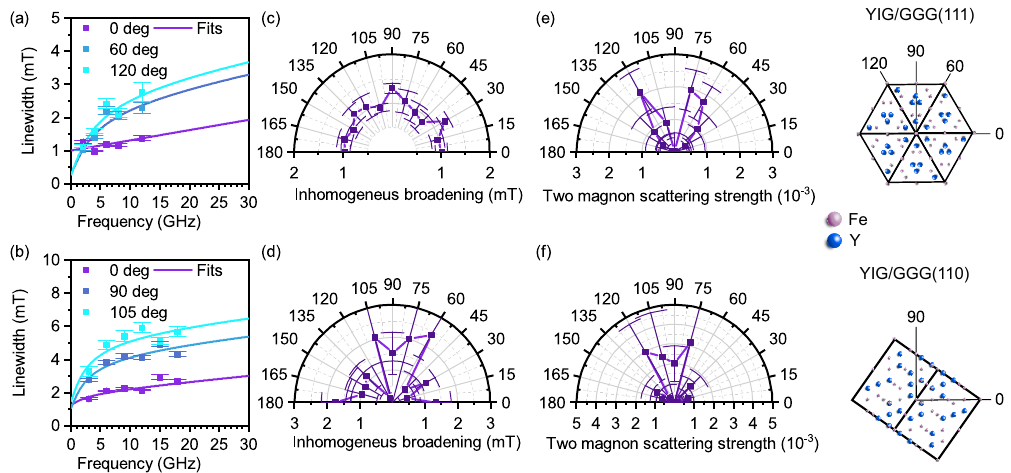}
\caption{\label{fig:Fig4}Angle-resolved FMR linewidth analysis.
(a,b) Frequency-dependent FMR linewidths measured at selected in-plane field angles for YIG/GGG(111) (a) and YIG/GGG(110) (b), together with fits using Eq.~\ref{eq:FMR_linewidth}.
(c,d) Extracted angle-dependent inhomogeneous broadening, $\mu_0\Delta H_0(\phi)$, for YIG/GGG(111) (c) and YIG/GGG(110) (d).
(e,f) Extracted two-magnon scattering strength, $\Gamma(\phi)$, for YIG/GGG(111) (e) and YIG/GGG(110) (f). 
The corresponding in-plane crystallographic directions for the two substrate orientations are shown on the right.
In all four panels the field is applied at $90^\circ$ to the depicted crystallographic direction.}
\end{figure*}

The FMR linewidth ($\mu_0 \Delta H$) was analyzed  assuming three contributions -- inhomogeneous broadening, Gilbert damping, and two-magnon scattering:
\begin{equation}
\begin{aligned}
\mu_0 \Delta H
&=
\frac{\alpha \omega}{\gamma}
+
\mu_0 \Delta H_0 \\
&\quad
+
\Gamma 
\arcsin\!\left[
\sqrt{
\frac{
\sqrt{\omega^{2}
+\left(\frac{\omega_\mathrm{M}}{2}\right)^{2}}
-\frac{\omega_\mathrm{M}}{2}
}{
\sqrt{\omega^{2}
+\left(\frac{\omega_\mathrm{M}}{2}\right)^{2}}
+\frac{\omega_\mathrm{M}}{2}
}
}
\right],
\end{aligned}
\label{eq:FMR_linewidth}
\end{equation}
where $\mu_0 \Delta H_0$ represents the inhomogeneous broadening contribution, $\alpha$ is the Gilbert damping parameter, $\Gamma\left( \phi \right)$ is two-magnon scattering strength,  and $\omega_\mathrm{M} = \mu_0\gamma M_s$. We show 3 selected angles and their respective fits for each sample on YIG(111) and YIG(110)(Fig.~\ref{fig:Fig4}a, b). The Gilbert damping was assumed to be the same for all angles $\phi$, while inhomogeneous broadening [$\mu_0 \Delta H_0 (\phi)$] and two-magnon scattering strength [$\Gamma(\phi)$] were allowed to vary.

The fit yields $\alpha_\mathrm{YIG/GGG(111)} = (8 \pm 6)\times10^{-4}$ and $\alpha_\mathrm{YIG/GGG(110)} = (9 \pm 6)\times10^{-4}$, indicating that the intrinsic damping is essentially the same for both orientations. Nevertheless, the differences in linewidths point to additional extrinsic contributions arising from inhomogeneous broadening and two-magnon scattering. For YIG/GGG(111), the inhomogeneous broadening, $\mu_0\Delta H_0(\phi)$, is essentially angle independent, whereas YIG/GGG(110) exhibits enhanced inhomogeneous broadening near $\phi = 90^\circ$ (Fig.~\ref{fig:Fig4}d). This behavior indicates that the reduced symmetry of YIG/GGG(110) gives rise to an anisotropic broadening contribution, while the higher in-plane symmetry of YIG/GGG(111) results in a nearly isotropic response \cite{medwal2021facet}.

In both samples, the two-magnon scattering strength $\Gamma(\phi)$ exhibits a pronounced angular anisotropy (Fig.~\ref{fig:Fig4}e,f). The maxima follow high-symmetry crystallographic directions of each respective substrate, indicating that the scattering is governed by crystallographically correlated perturbations of the effective magnetic field \cite{medwal2021facet}. Within the framework developed above, a maximum in $\Gamma(\phi)$ corresponds to efficient coupling between the uniform FMR mode and degenerate spin-wave states, which is expected when the dominant perturbation is elongated perpendicular to the field direction at which the maximum occurs. Such perturbations may originate from anisotropic surface morphology including elongated steps or line defects, or from low-angle grain boundaries. YIG/GGG(110) exhibits a  two-fold symmetry, with a maximum scattering associated with the $[1\bar{1}\bar{1}]$ direction. Substrate miscut can superpose a two-fold symmetry on the crystal symmetry, which may explain the inequivalent response in the YIG/GGG(111) sample between the $0^\circ$ direction and the two maxima observed along the 60\,deg and 120\,deg directions.  While the origin of the effective field perturbations in these YIG films remains to be determined, this work suggests that the relationship between perturbation geometry and two-magnon scattering can be  quantitatively characterized, for example by patterning topographic or other features into the YIG films with well defined length scale and orientation. To illustrate this, a micromagnetic model of the scattering potential provided by a 6\,nm high step in a YIG film is presented in the End Matter, showing an angular dependence of linewidth in agreement with the theoretical model.  

In conclusion, we have introduced a framework for describing two-magnon scattering based on the Lippmann--Schwinger equation. Within the Born approximation, the scattering efficiency is governed not only by the density and strength of perturbations in the effective field (i.e. 'defects'), but also by the spatial symmetry of the defect potential and its reciprocal-space overlap with the degenerate spin-wave states. Directional defects, such as elongated structural inhomogeneities, therefore produce an intrinsically anisotropic two-magnon contribution whose strength can be controlled by the orientation between the magnetic field and the defect landscape. We applied this framework to YIG films grown on GGG(111) and GGG(110), for which the FMR linewidth exhibits a pronounced in-plane angular dependence. The anisotropic linewidth enhancement is consistent with two-magnon scattering from  defect potentials that are elongated along crystallographic directions.  The framework presented here  provides a pathway to identify, suppress, or engineer extrinsic linewidth contributions. Although demonstrated here for scattering from the uniform FMR mode, the framework is general and applies equally to propagating spin waves. This understanding is essential for minimizing losses in magnonic devices and for designing geometries in which defect-induced scattering is either avoided or deliberately exploited.

\begin{acknowledgments}
The authors acknowledge the MIT Office of Research Computing and Data
for providing high-performance computing resources and the use of shared facilities at MIT.nano. The authors acknowledge support from the U.S. National Science
Foundation under Awards No.~ECCS-2232830, No.~OMA-2326754, and
No.~ECCS-2328839. O.W. was supported by Horizon Europe - MSCA grant agreement No. 101211677, project FeriMag. 
\end{acknowledgments}

\noindent\textit{Data availability ---}
The experimental and simulated data underlying the figures,
together with the code used to generate the numerical results in
Fig.~1, are openly available in Zenodo~\cite{marquez_chavez_2026}.

\bibliography{apssamp}

\appendix
\section{End Matter}
\subsection{Deposition of the YIG films}
The films were deposited at \(900^\circ\mathrm{C}\) in an oxygen atmosphere of 25\,mTorr, with a base pressure of $1\times 10^{-5}$\,mTorr, using a 248\,nm KrF excimer laser with 400\,mJ energy and a repetition rate of 10\,Hz. The deposition rates were determined by x-ray reflectometry to be about 70\,nm for YIG/GGG(111) and 24\,nm for YIG/GGG(110). No in-situ postannealing was performed, and the films were cooled down to \(200^\circ\mathrm{C}\) in the  same oxygen pressure. Atomic force microscopy showed root-mean-square (RMS) surface roughness values of 0.106 nm for the YIG/GGG(111) film and 0.206 nm for the YIG/GGG(110) film.

\subsection{Micromagnetic simulations}
\begin{figure}
\centering
\includegraphics{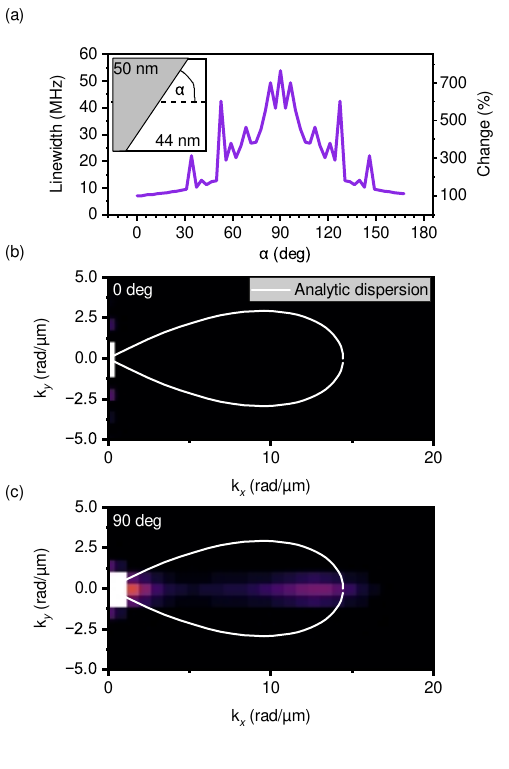}
\caption{\label{fig:sim} Micromagnetic simulations of two-magnon scattering.
(a) Simulated FMR linewidth as a function of the angle between the topographical step in the YIG and the applied magnetic field. Only the range $0^\circ$--$90^\circ$ was simulated; the response from $90^\circ$ to $180^\circ$ was obtained by symmetry. 
(b,c) Reciprocal-space distribution of the excited states for step orientations of $0^\circ$ (b) and $90^\circ$ (c), showing suppressed and enhanced occupation of finite-$k$ states, respectively.}
\end{figure}

To test the predictions of the theoretical framework developed here, we performed micromagnetic simulations using MuMax3. The simulated film had lateral dimensions of approximately $8 \times 8 \mu\mathrm{m}^2$ and a thickness of $50$ nm, with a cell size of approximately $8 \times 8 \times 6$\ nm$^3$. The defect was modeled as a one-cell-high thickness step, corresponding to a height variation of $6$ nm. This step extended across the entire simulated film, divided the sample into two equal halves, and  passed through the center of the simulation area. The angle between the step and the applied magnetic field was then  varied.

We first analyzed the simulated FMR linewidth as a function of the defect orientation, as shown in Fig.~\ref{fig:sim}a. In agreement with the Lippmann--Schwinger framework, the linewidth is minimum when the step is parallel to the applied field, i.e., at $0^\circ$. As the angle increases, the linewidth gradually increases and reaches its maximum when the step is perpendicular to the field. This behavior is further corroborated by the reciprocal-space distribution of the excited states. For a step oriented at $0^\circ$ (Fig.~\ref{fig:sim}b), no appreciable occupation of $k \neq 0$ states is observed, indicating suppressed two-magnon scattering. In contrast, for a step oriented at $90^\circ$ (Fig.~\ref{fig:sim}c), pronounced occupation of finite-$k$ states appears in the backward-volume geometry, confirming efficient scattering from the uniform FMR mode into degenerate spin-wave states.

\end{document}